\documentclass[a4paper]{article}

\usepackage{INTERSPEECH2022}
\usepackage{multicol}
\usepackage{multirow}

\title{Multi-stage Progressive Compression of Conformer Transducer for On-device Speech Recognition}

\name{Jash Rathod$^1$, Nauman Dawalatabad$^2$, Shatrughan Singh$^1$, Dhananjaya Gowda$^1$ \thanks{$^2$Work done while at Samsung Research.}}
\address{
  $^1$Samsung Research\\
  $^2$Massachusetts Institute of Technology}
\email{\{jash.rathod, shatrughan.s, d.gowda\}@samsung.com, nauman@csail.mit.edu}

\begin{document}

\maketitle

\begin{abstract}

The smaller memory bandwidth in smart devices prompts development of smaller Automatic Speech Recognition (ASR) models. To obtain a smaller model, one can employ the model compression techniques. Knowledge distillation (KD) is a popular model compression approach that has shown to achieve smaller model size with relatively lesser degradation in the model performance. In this approach, knowledge is distilled from a trained large size teacher model to a smaller size student model. Also, the transducer based models have recently shown to perform well for on-device streaming ASR task, while the conformer models are efficient in handling long term dependencies. Hence in this work we employ a streaming transducer architecture with conformer as the encoder. We propose a multi-stage progressive approach to compress the conformer transducer model using KD.
We progressively update our teacher model with the distilled student model in a multi-stage setup. 
On standard LibriSpeech dataset, our experimental results have successfully achieved compression rates greater than 60\% without significant degradation in the performance compared to the larger teacher model.

\end{abstract}


\section{Introduction}
 
Automatic Speech Recognition (ASR) on smart devices traditionally involved sending data to a server, performing speech recognition, and returning the results to the device \cite{he2019streaming}. While this approach helps us leverage more computing power, it incurs increased latency. It would be better if this could be done on-device itself. This demands models to work with a smaller memory footprint \cite{sainath2020streaming}.
Recently, there has been great adoption of End-to-End (E2E)  ASR systems in smart devices. \cite{he2019streaming,sainath2019two,ganesh_ondevice,samsung_streaming, garg2019asru, infra_paper, sr_vd_is2020}. 
Also, for many use cases, it is desirable to use a streaming model as a real-time solution is expected \cite{yu2020dual}. Thus, it is crucial that our on-device ASR model adhere to these constraints and at the same time not compromise on the performance.
 
Recurrent Neural Network Transducer (RNN-T) is a widely used sequence-to-sequence streaming model for ASR~\cite{graves2012sequence}. It is a low-latency on-device solution that performs well when used with edge devices.
The RNN-T model comprises a transcription network, a prediction network, and a joint network.
The transcription network encodes the input audio using Long Short Term Memory (LSTM) cells \cite{sak2014long, sherstinsky2020fundamentals}.
Recently in \cite{gulati2020conformer}, authors show that the conformer model efficiently learns local as well as global interactions.
Hence, the transcription network of the RNN-T, which originally uses unidirectional LSTM blocks, can be replaced with conformer \cite{gulati2020conformer} block forming a conformer-transducer model. 
 
Smaller models generally observe degradation in their performance as they cannot learn as effectively as their larger counterparts.
This is due to the smaller model capacity. One of the methods to obtain the smaller models is by using model compression techniques \cite{cheng2018model, wu2016quantized, yu2017compressing}. Knowledge distillation (KD) \cite{hinton2015distilling, chebotar2016distilling, Gou_2021} is a popular model compression technique that has recently been successful across various domains like natural language processing \cite{turc2019well, kim2016sequence, sanh2019distilbert}, computer vision \cite{liu2019structured, chen2017learning}, and speech recognition \cite{ren2020fastspeech, dawalatabad2021two}. In this technique, knowledge is distilled from a larger model (teacher) to a smaller model (student). 
More recently, KD has also been shown to give a high compression rate with RNN-T based two-pass ASR model~\cite{dawalatabad2021two}.

Using an even larger teacher model to distill a student model can be one avenue to improve the performance of a student, leveraging the rich knowledge learned by the teacher. However, our empirical results 
and recent works \cite{mirzadeh2020improved} suggest that a higher gap in sizes tends to impair the model performance, even though the larger teacher performs significantly better than a smaller teacher. This led us to the idea of bridging this gap by training a student model using a multi-stage approach.
In this work, we propose multi-stage progressive compression for the conformer-transducer model.  
While \cite{mirzadeh2020improved} discussed this idea for CNNs, our work is the first one to introduce this technique for ASR systems, to the best of our knowledge.
In Stage 1, we use a large teacher model and compress it using KD such that it does not degrade significantly. 
In the next stage, the student model obtained in Stage 1 serves as the teacher model for the current stage and we distill a new student model for this stage. This process is repeated multiple times till a student model with the desirable size is achieved. 
The final student model obtained can then be used for on-device ASR. 
We perform detailed studies by comparing the results of the student models in every stage with models of the same configuration trained from scratch. We found that the student model always outperforms the model of the same size trained from scratch.

\vspace{-0.1cm}

\section{Methodology}
\label{sec:methodology}

\begin{figure}[t]
  \centering
  \includegraphics[width=0.8\linewidth]{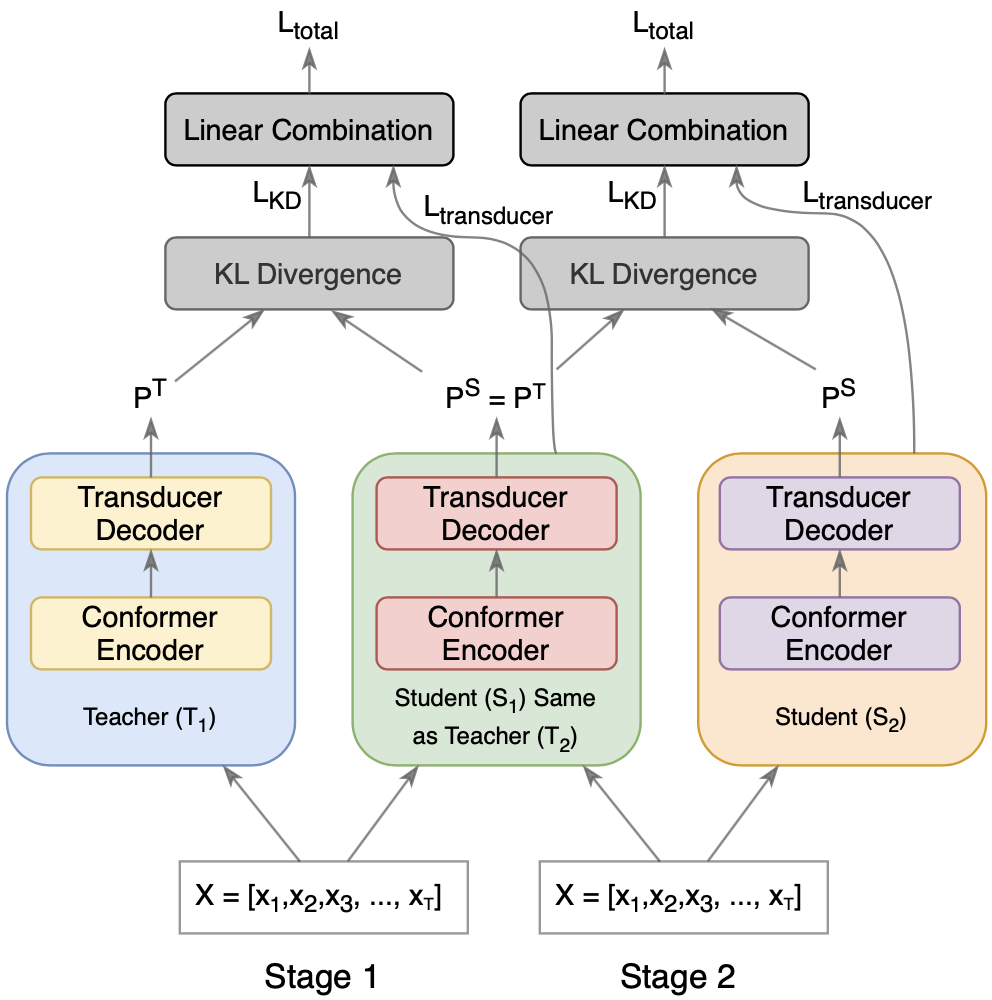}
  \caption{Multi-stage progressive compression of conformer transducer model. The student model distilled from teacher in the first stage, is used as the teacher model for the next stage. This cascaded approach can be continued for $N$ stages.}
  \label{fig:multistage_methodology}
\end{figure}

In this section, we first describe the modules in the conformer transducer model used in this paper.
We then describe the multi-stage progressive KD approach for compression of the conformer-transducer model as shown in Figure \ref{fig:multistage_methodology}.

\subsection{Conformer}
\label{ssec:conformer}
 
The conformer block \cite{gulati2020conformer}  includes a feed-forward module, followed by a convolutional network, a multi-headed self-attention \cite{vaswani2017attention} module, and finally another feed-forward module. 
The advantage of using this architecture is that the attention module helps in learning long-range global interactions well and the convolutional module exploits the local features. 

Assume input given to the conformer block $i$ is $x_i$ and its output to be $y_i$. Mathematically, we can represent the conformer block as follows:

\begin{equation}
    \begin{gathered}
        x_{i}^{'} = x_i + FF(x_i)\\
        x_{i}^{''} = x_{i}^{'} + Conv(x_{i}^{'})\\
        x_{i}^{'''} = x_{i}^{''} + MA(x_{i}^{''})\\
        y_i = LayerNorm(x_{i}^{'''} + FF(x_{i}^{'''})) \label{eq:conformer}
    \end{gathered}
\end{equation}

\noindent where $FF$ refers to the feed-forward module, $Conv$ refers to the convolutional module, and $MA$ refers to the multi-headed self-attention module.

The output obtained from the conformer blocks are the higher-level representations that are fed to the joint network of the transducer.

\subsection{Transducer}
\label{ssec:transducer}

The RNN-Transducer is a streaming sequence-to-sequence model that has an encoder-decoder architecture \cite{graves2012sequence}. It includes a transcription, a prediction, and a joint network. The transcription network takes audio as the input. The prediction network, which is an autoregressive network, takes the previous output as input. They produce higher-level representations that are fed to the joint network (feed-forward network) to predict the posterior probability of the current token. In this work, we use the conformer blocks instead of LSTMs as the transcription network.  

\subsection{Knowledge distillation for conformer transducer}
\label{ssec:kd}

The technique of knowledge distillation aims at harnessing the rich information, that a large model learns, into a smaller model \cite{hinton2015distilling}. The advantage of this approach is that we obtain a smaller model that performs well and due to its size, is a better choice for being deployed to edge devices. In knowledge distillation, a teacher-student architecture is used. The output probability distribution (soft targets) represents the knowledge of the teacher model. To enable a student to make use of this, we minimize the KL-Divergence between the soft targets of the teacher and the student model. 
In our case, the KL-Divergence between the output of the joint network of the teacher and that of the student is minimized. 
If $k$ represents the output label and $(t,u)$ is the lattice node, the KL-Divergence loss can be represented as follows:

\begin{equation}
    L_{KD} = \sum_{t,u}\sum_{k} P^{T}(k|t,u)\ln\frac{P^{T}(k|t,u)}{P^{S}(k|t,u)}
\end{equation}

Let $\alpha$ be the distillation loss weight. Then, the total loss for training a conformer transducer student model from a conformer transducer teacher model is given as follows:

\begin{equation}
    L_{total} = (1-\alpha)L_{transducer} + \alpha L_{KD}
\end{equation}

\subsection{Multi-stage progressive knowledge distillation}
\label{sec:proposed}

As discussed earlier, the use of a larger and better-performing teacher model should be beneficial. Yet, if the compression rate increases beyond a limit, existing works \cite{mirzadeh2020improved} and even our empirical results (discussed in Section~\ref{ssec:ablation}) indicate a sharp degradation. It is likely that the complex relations learned by the larger teacher model are too difficult for a smaller student model to learn with low model capacity. As a model compressed using KD outperforms a model of the same size trained from scratch, it is desirable to use a KD-compressed model.
Hence, we compress a large streaming teacher model using knowledge distillation in multiple stages to attain a better-performing streaming student model. Figure \ref{fig:multistage_methodology} gives a schematic representation of the proposed solution. In the first stage, we compress the teacher model at a compression rate that does not hurt its performance beyond a tolerable bandwidth. We obtain an intermediate high performing student model which has the potential to be compressed further. In the next stage, we compress the student model from the previous stage (which serves as the teacher for the next stage) to attain a smaller student model. This process can be performed progressively for a number of stages till the required model size is achieved. The intuition behind this is to not lose the complex interactions learned by the larger teacher by adopting a progressive compression approach, as direct compression does not allow the student to learn these interactions. 

\begin{table*}[t]
  \centering
  \caption{Multi-stage progressive compression experiment results, including number of model parameters (in million) (rounded off to whole number), compression (Comp) percentage (rounded off to whole number), WERs and SERs for test-clean and test-other LibriSpeech datasets. In experiment 1, we compress 128M params teacher model to 46M params student model in three stages. In experiment 2, we compress 128M params teacher model to 24M params student model in two stages.}
  \resizebox{0.95\textwidth}{!}{
  \begin{tabular}{c|c|c|c|c|c|c|c|c|c}
    \toprule
    \multirow{2}{*}{\textbf{Experiment}} & \multirow{2}{*}{\textbf{Stage}} & \multirow{2}{*}{\textbf{Model}} & \textbf{Parameters} & \textbf{\% Comp w.r.t.} & \textbf{\% Comp w.r.t.} & \multicolumn{2}{c|}{\textbf{test-clean}} & \multicolumn{2}{c}{\textbf{test-other}} \\
    \cline{7-10}
    & & & \textbf{(in million)} & \textbf{teacher (\%)} & \textbf{$T_1$ (\%)} & \textbf{WER} & \textbf{SER} & \textbf{WER} & \textbf{SER} \\ 
    \midrule
    \midrule
    \multirow{11}{*}{Experiment 1} & \multirow{3}{*}{Stage 1} & $T_1$ & 128 & - & - & 5.54 & 52.71 & 14.99 & 77.65 \\
    & & $B_1$ & 80 & - & 38 & 5.98 & 54.81 & 16.12 & 79.35 \\
    & & $S_1$ (Student from $T_1$) & 80 & 38 & 38 & 5.47 & 51.76 & 14.96 & 77.17 \\
    \cline{2-10}
    & \multirow{4}{*}{Stage 2} & $T_2 = S_1$ & 80 & 38 & 38 & 5.47 & 51.76 & 14.96 & 77.17 \\
    & & $B_2$ & 62 & - & 52 & 6.19 & 55.99 & 16.66 & 79.38 \\
    & & $S_2$ (Student from $T_1$) & 62 & 52 & 52 & 5.67 & 52.91 & 15.01 & 77.82 \\
    & & $S_3$ (Student from $T_2$) & 62 & 23 & 52 & 5.48 & 52.18 & 14.85 & 77.17 \\
    \cline{2-10}
    & \multirow{4}{*}{Stage 3} & $T_3 = S_3$ & 62 & 23 & 52 & 5.48 & 52.18 & 14.85 & 77.17 \\
    & & $B_3$ & 46 & - & 64 & 6.65 & 58.89 & 17.03 & 79.96 \\
    & & $S_4$ (Student from $T_1$) & 46 & 64 & 64 & 6.20 & 56.04 & 16.38 & 78.9 \\
    & & $S_5$ (Student from $T_3$) & 46 & 25 & 64 & \textbf{6.08} & \textbf{55.53} & \textbf{16.11} & \textbf{78.73} \\
    \midrule \midrule 
    \multirow{7}{*}{Experiment 2} & \multirow{3}{*}{Stage 1} & $T_1$ & 128 & - & - & 5.54 & 52.71 & 14.99 & 77.65 \\
    & & $B_4$ & 57 & - & 55 & 6.32 & 56.3 & 16.72 & 79.72 \\
    & & $S_6$ (Student from $T_1$) & 57 & 55 & 55 & 5.76 & 53.36 & 15.44 & 77.95 \\
    \cline{2-10}
    & \multirow{4}{*}{Stage 2} & $T_4 = S_6$ & 57 & 55 & 55 & 5.76 & 53.36 & 15.44 & 77.95 \\
    & & $B_5$ & 24 & - & 81 & 8.17 & 64.24 & 20.17 & 83.63 \\
    & & $S_7$ (Student from $T_1$) & 24 & 81 & 81 & 7.33 & 60.99 & 18.88 & 83.12 \\
    & & $S_8$ (Student from $T_4$) & 24 & 58 & 81 & \textbf{7.21} & \textbf{60.11} & \textbf{18.59} & \textbf{83.06} \\
    \bottomrule
  \end{tabular}
  }
  \label{tab:multistage}
\end{table*}

\vspace{-0.1cm}
\section {Experimental Setup}
\label{sec:expsetup}

\subsection{Data and features}

The dataset used for this work is the standard LibriSpeech dataset \cite{panayotov2015librispeech} which contains 960 hours of training data. Models are evaluated on the `test-clean' set (5.4 hours; contains 2620 utterances) and `test-other' set (5.1 hours; contains 2939 utterances). The features used are 40-dimensional Mel Frequency Cepstral Coefficients (MFCCs) \cite{wanli2013research}  which are calculated from a 25~msec window for every 10~msec shift and are passed as an input to the transcription network. The target vocab consists of 10026 Byte Pair Encoding (BPE) \cite{sennrich2015neural} units. We use SpecAugment \cite{park2019specaugment} technique for data augmentation during training. The beam size of 8 is used for decoding.

\subsection{Transducer model and KD specifications} 

The specifications of each of the teacher and student models used can be found in Table \ref{tab:multistage}. Other attributes remain the same across all models which are discussed next in this section. 
The entire training pipeline is developed using Tensorflow 2 \cite{tensorflow2015-whitepaper}. A dropout \cite{srivastava2014dropout} of 0.1 is used succeeding every layer other than the first one. Max-pooling of 2 for the first three layers (and an overall factor of 8) is used which helps in shortening the training time and latency of the model.  
To reduce the number of parameters and regularize our model, the same neural network serves as the joint network and the embedding layer in the prediction network.
We use a learning rate represented by $lr$ of $0.0001$ and decay it using the decay function as follows:

\vspace{-0.3cm}
\begin{equation}
    \begin{gathered}
        \lambda = \frac{step}{warmup}\\
        \gamma = \frac{step - warmup}{decay\_steps}\\
        lr= 
        \begin{cases}
            lr * \lambda^3 + 10^{-7},    & \text{if } step < warmup\\
            min(lr, 5 * lr * 0.9^\gamma), & \text{otherwise}
        \end{cases}
    \end{gathered}
\end{equation}

\noindent where $step$ represents current training time step, $warmup$ is set to be $10000$, and $decay\_steps$ is set to be $40000$. 

As for knowledge distillation, our preliminary experiments show that a distillation loss weight of $0.02$ gives the best results. For all our experiments, a temperature \cite{hinton2015distilling} of $1.0$ has been used for both, the teacher and the student models.

\subsection{Notations used}

In this paper, we denote a teacher model with $T$ and a student model with $S$. For a detailed comparison, we have also trained baselines models  $B$ with the same configuration as the student models, trained from scratch.   
Thus, $T$ and $B$ are models trained without knowledge distillation. 
A subscript acts as identification between different models of the same type. 

Table \ref{tab:multistage} includes the  experiments performed, their model sizes in number of parameters in million (M), the compression percentages (\%), and the evaluation metrics - Word Error Rate (WER) and Sentence Error Rate (SER). The  $T_i = S_j$ indicates that $S_j$ now serves as $T_i$ and both are the same models. 

\vspace{-0.1cm}
\section {Results and Discussion}
\label{sec:results}

In this section, we exhibit the efficacy of our proposed multi-stage progressive compression approach. We present results on varying compression rates and their performance as discussed in Section \ref{ssec:kd}. All models in these experiments are streaming models. Each model was trained on two NVIDIA A100 GPUs for 80 epochs. Also, all the results are without the use of a Language Model (LM).  
For every stage, compression is achieved by reduction in encoder and decoder dimensions and/or number of layers.

\subsection{Experiment 1 - compression rate of under 50\% in each stage}

\begin{figure}[t]
  \centering
  \includegraphics[width=\linewidth]{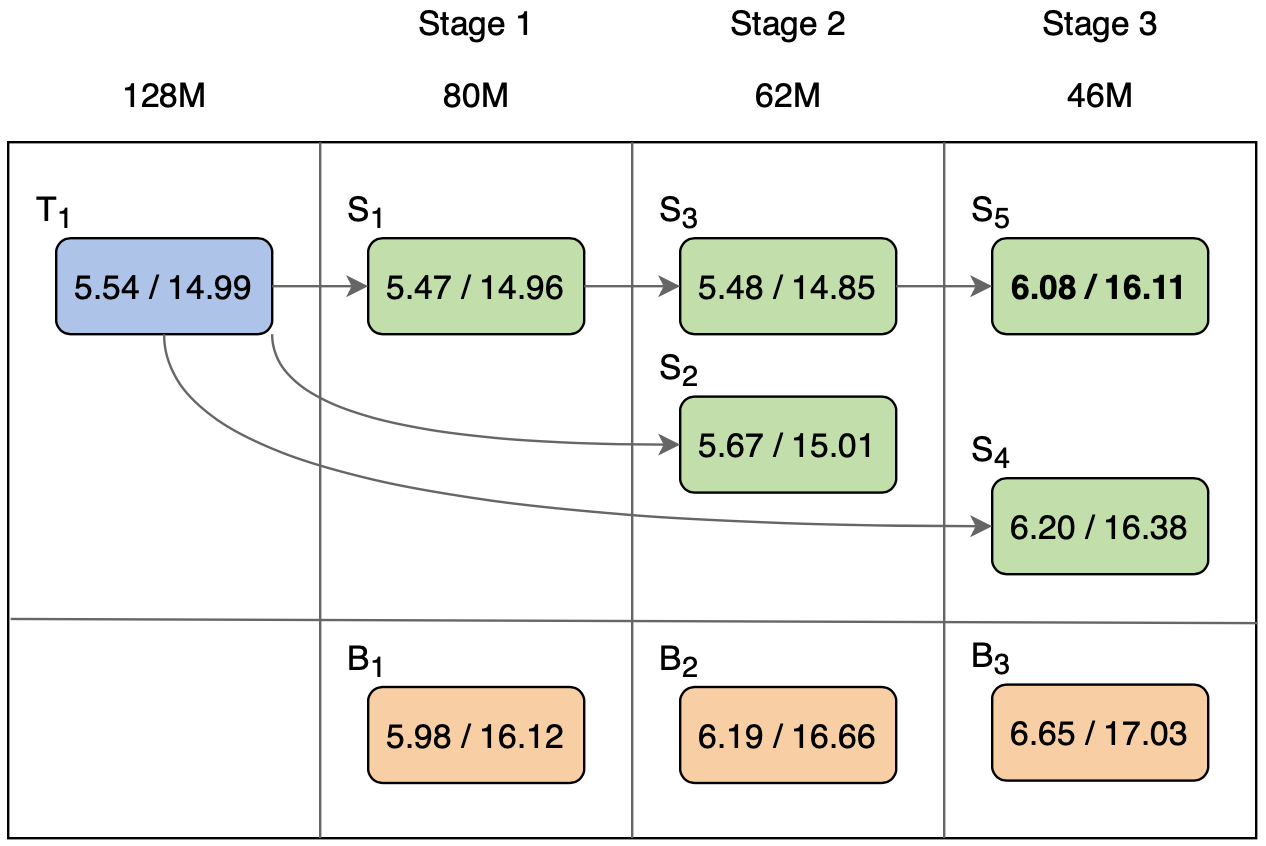}
  \caption{Results for multi-stage KD compression of conformer transducer  Experiment 1 (compression rate of under 50\% in each stage) on test-clean / test-other sets. $T_i$, $S_j$, and $B_k$ represent teacher, student, and baseline models respectively.}
  \label{fig:results_exp1}
\end{figure}

Table \ref{tab:multistage} and Figure~\ref{fig:results_exp1} highlight results of multi-stage progressive compression approach where we perform two sets of experiments. In experiment 1, we compress a 128M parameters teacher model to a 46M parameters student model in three stages progressively.  
The 128M teacher model has 16 encoder layers with 512 dimensions, 4 attention heads, and 2 decoder layers with 1024 dimensions.
In each stage, we have kept the compression rate below 50\% compared to the teacher model size in that stage. It can be observed that in Stage 1, we obtained the student model $S_1$ (distilled from $T_1$) with better WER than the teacher and 38\% smaller in size. For comparison, we also trained a baseline model $B_1$ (same configuration model but trained from scratch without KD). As expected, $B_1$ is found to be worse than the student model $S_1$. 

In Stage 2 of multi-stage progressive KD, the $S_1$ model is used as the teacher model $T_2$. Even here, we observe that our student model $S_3$ has a WER almost the same as that of the teacher while being  23\% smaller than $T_2$.  
Again, clearly student model $S_3$ outperforms $B_2$ and $S_2$ (distilled from $T_1$).
It is worth noticing that, compared to the $T_1$, our student model $S_3$ trained through the proposed progressive KD has been compressed by 52\% with no loss in WER performance. 

For Stage 3, $S_3$ model now acts as the teacher model $T_3$. We went a step ahead and compressed the model by 25\% which brings the model size to 46M parameters. 
This time we observe approximately 11\% and 8.5\% relative degradation in WER on test-clean and test-other respectively. 
This minor degradation in WER is reasonable 
given a high compression rate of 64\% with respect to original teacher model $T_1$. 
Also, when compared to the baseline $B_3$ model, the student $S_4$ and $S_5$ performs consistently better again. 
$S_5$ performing better than $S_4$ shows that the student model obtained using our multi-stage progressive compression technique produces the most superior results.
Compared to original teacher model $T_1$, the minor relative degradations of 10\% and 7.5\%  in WERs on test-clean and test-other datasets are observed. 
At the same time, a significantly high compression rate of 64\% compared to the teacher model is achieved.

\subsection{Experiment 2 - compression rate of over 50\% in each stage}

\begin{figure}[t]
  \centering
  \includegraphics[width=\linewidth]{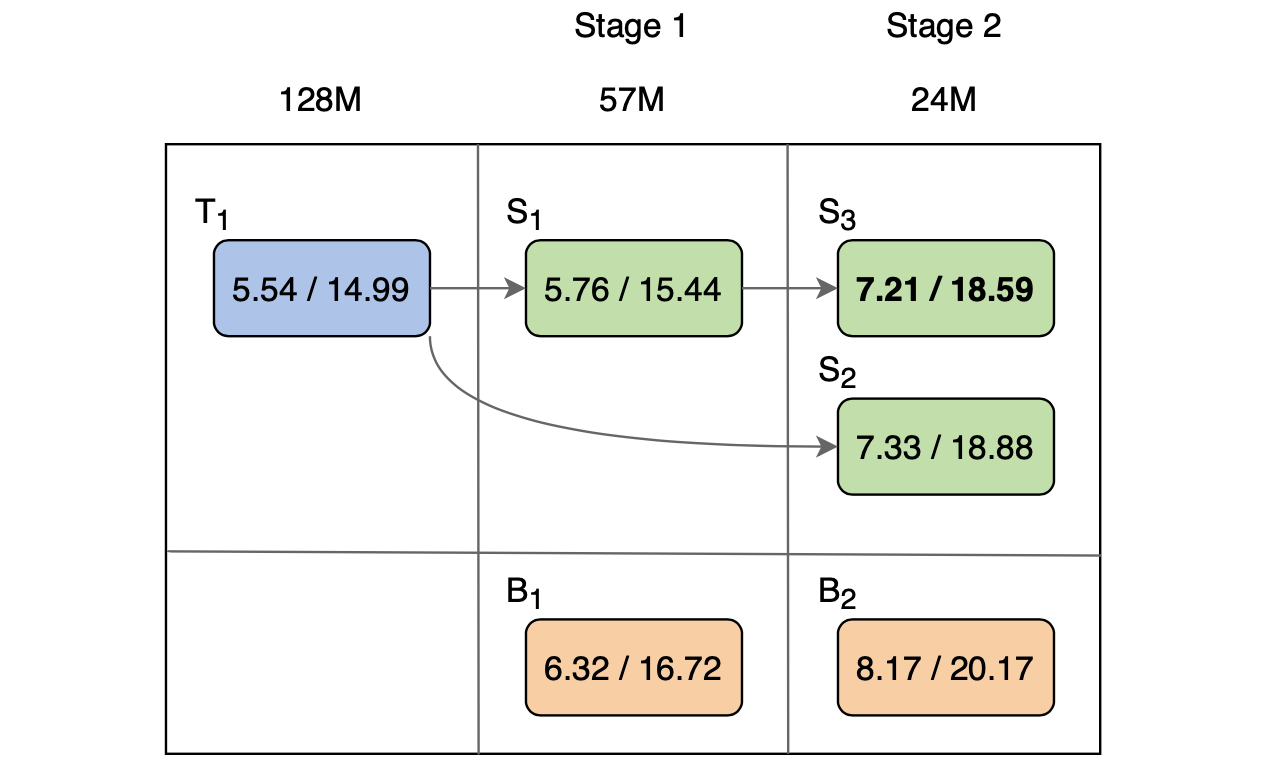}
  \caption{Results for multi-stage KD compression of conformer transducer  Experiment 2 (compression rate of over 50\% in each stage) on test-clean / test-other sets. $T_i$, $S_j$, and $B_k$ represent teacher, student, and baseline models respectively.}
  \label{fig:results_exp2}
\end{figure}
 
In order to experiment with still higher compression rates (beyond 55\%), we performed experiment 2 which uses two stages. The results of this set of experiments are shown in bottom row of Table \ref{tab:multistage} and in Figure~\ref{fig:results_exp2}. 
For Stage 1, we use the same 128M parameters teacher $T_1$. We distill a student model $S_6$ from $T_1$ with a compression rate of 55\% and observe minor degradation of approximately  4\% and 3\% (relative) in WER on the test-clean and test-other datasets respectively. Also, we observe that the performance of the baseline $B_4$ is significantly worse.

For Stage 2, we compress the teacher model $T_4$ (i.e. $S_6$) by 58\% to obtain a model with only 24M parameters.  
While $S_7$ performs better than the $B_5$, it is marginally worse than $S_8$. We get a degradation in WER of about 24\% (relative) on both test-clean and test-other. This degradation in WER is expected given a high compression rate of 81\% with respect to $T_1$.

\subsection{Ablation study on single stage KD}
\label{ssec:ablation}

In this section, we demonstrate the effect of compression rates higher than 65\% using only a single stage. From results in Table \ref{tab:ablation}, we can observe that generally as the compression increase, the WERs tend to degrade significantly. $S_9$ and $S_{10}$ are identical except for a higher decoder compression rate used in the latter. $S_{11}$ and $S_{12}$ are also identical except for a higher decoder compression rate used in the latter and both have higher encoder compression rate than $S_9$ and $S_{10}$. Considering sets ($S_{9}$, $S_{13}$) and ($S_{12}$, $S_{14}$), each models in their respective sets are identical except for higher encoder compression rates used in the latter.  
Based on our studies, we have been able to draw some insights as in~\cite{botros2021tied}. Firstly, compression in decoder doesn't affect the performance of the models as much. And secondly, even a slightly higher compression in the encoder parameters (number of layers and input sizes) affects the model performance.  

\begin{table}[t]
  \centering
  
  \caption{Higher compression rates on 128M param teacher model using single stage knowledge distillation}
  \begin{tabular}{c|c|c|c|c|c}
    \toprule
    \textbf{Model} & \textbf{M Params} & \multicolumn{2}{c|}{\textbf{test-clean}} & \multicolumn{2}{c}{\textbf{test-other}} \\\cline{3-6}
    & \textbf{(\% Comp)} & \textbf{WER} & \textbf{SER} & \textbf{WER} & \textbf{SER} \\ 
    \midrule
    \midrule
    $T_1$ & 128 (-) & 5.54 & 52.71 & 14.99 & 77.65 \\
    \hline
    $S_9$ & 41 (68\%) & 6.21 & 56.34 & 16.53 & 79.41 \\
    $S_{10}$ & 37 (71\%) & 6.25 & 56.26 & 16.63 & 79.72 \\
    $S_{11}$ & 30 (76\%) & 7.09 & 60.08 & 18.05 & 81.59 \\
    $S_{12}$ & 27 (79\%) & 6.96 & 59.25 & 18.28 & 82.44 \\
    \hline
    $S_{13}$ & 29 (77\%) & 7.81 & 62.75 & 20.07 & 84.38 \\
    $S_{14}$ & 19 (85\%) & 9.32 & 68.21 & 22.55 & 87.24 \\
    \bottomrule
  \end{tabular}
  \label{tab:ablation}
\end{table}

\vspace{-0.2cm}
\section {Conclusion}
\label{sec:conclusion}

In this work, we propose a multi-stage progressive approach to compress conformer transducer model using knowledge distillation.
The proposed approach begins with a 128M parameters streaming teacher model and distills a smaller streaming model without any
degradation in performance of the model. In the next stage, this student model is used as the teacher model to obtain a smaller model for KD in this stage. Using this cascaded approach, we obtain two models: a 46M parameter model with a compression of 64\% in three stages, and a 24M parameter model with a compression of 81\% in two stages. The former shows a minor degradation, yielding 10\% and 7.5\% relative in WERs on LibriSpeech's test-clean and test-other datasets respectively. These compression rates are good enough for medium-end mobile devices.
Further, we also demonstrate the effects of single stage compression of a large teacher and share our insights.
In future, 
we aim to focus specifically on the encoder compression as it makes up a sizable portion of the model size and has a significant impact on performance.

\bibliographystyle{IEEEtran}

\bibliography{main}

\end{document}